\documentclass[11pt,letterpaper]{article}
\usepackage[left=1.4in, right=1.4in, top=1.2in, bottom=1.2in]{geometry}

\usepackage{authblk}
\usepackage{amsmath}
\usepackage{amscd}
\usepackage{amsthm}
\usepackage{amssymb} \usepackage{latexsym}
\usepackage{eufrak}
\usepackage{euscript}
\usepackage{epsfig}
\usepackage{graphics}
\usepackage{array} 
\usepackage{enumerate}
\usepackage{hyperref}
 
\usepackage{pictexwd,dcpic}

\usepackage[capitalize]{cleveref}

\usepackage{parskip}

\usepackage{color}

\newcommand{\bel}[1]{\begin{equation}\label{#1}}

\newcommand{\be}{\begin{equation}}

\newcommand{\ba}{\begin{eqnarray}}
\newcommand{\ea}{\end{eqnarray}}
\newcommand{\rf}[1]{(\ref{#1})}

\newcommand{\qe}{\end{equation}}

\renewcommand{\theequation}{\arabic{equation}}
\newtheorem{thesis}{Thesis}
\newcommand{\btl}[1]{\begin{thesis}\label{#1}}
\newcommand{\et}{\end{thesis}}

\theoremstyle{theorem}

\theoremstyle{corollary}

\theoremstyle{lemma}
\theoremstyle{definition}

\theoremstyle{proof}

\theoremstyle{remark}

\makeatletter
\newenvironment{newnumbering}[1][Roman]
 {
  \def\jr@counter{#1}%
  \jr@setup@numbering{#1}{1}%
 }
 {%
  \setcounter{jr@\jr@counter @equation}{\value{equation}}%
  \setcounter{equation}{\value{jr@equation@save}}%
  \ignorespacesafterend
 }
\newcounter{jr@equation@save}
\newcommand{\jr@setup@numbering}[2]{%
  \@ifundefined{c@jr@#1@equation}{\newcounter{jr@#1@equation}}{}%
  \setcounter{jr@equation@save}{\value{equation}}%
  \setcounter{equation}{%
    \ifnum#2>0
      \value{jr@#1@equation}%
    \else
      0%
    \fi
  }%
  \renewcommand{\theequation}{\csname#1\endcsname{equation}}%
}
\makeatother
\usepackage{tikz-cd} 

\usepackage{tikz}

\usepackage{todonotes}

\title{\textit{Game Manipulators} - the Strategic Implications\\ of Binding Contracts}

\author[1,2]{María Alejandra Ramírez}
\author[3]{Yoav Kolumbus}
\author[4]{Rosemarie Nagel} 
\author[5,6,7,8]{David Wolpert}
\author[1,5,9]{J\"urgen Jost}

\affil[1]{\small{Max Planck Institute for Mathematics in the Sciences, Leipzig, Germany}}

\affil[2]{Max Planck Institute for Evolutionary Biology, Plön, Germany}

\affil[3]{Cornell University, USA}

\affil[4]{ICREA, Department of Economics, Universitat Pompeu Fabra, Barcelona School of Economics,\, 
Barcelona, Spain}

\affil[5]{Santa Fe Institute for the Sciences of Complexity, Santa Fe, NM, USA}

\affil[6]{Complexity Science Hub, Vienna, Austria}

\affil[7]{Arizona State University, Tempe, AZ, USA}

\affil[8]{International Center for Theoretical Physics, Trieste, Italy}

\affil[9]{Center for Scalable Data Analytics and Artificial Intelligence, Leipzig University, Germany}

\date{}

\begin{document}

\maketitle

\begin{abstract}
Commitment devices are powerful tools that can influence and incentivise certain behaviours by linking them to rewards or punishments. These devices are particularly useful in decision-making, as they can steer individuals towards specific choices. In the field of game theory, commitment devices can alter a player's payoff matrix, ultimately changing the game's Nash equilibria. Interestingly, agents, whom we term \textit{game manipulators} and who can be external to the original game, can leverage such devices to extract fees from players by making them contingent offers that modify the payoffs of their actions. This can result in a different Nash equilibrium with potentially lower payoffs for the players compared to the original game. For this scheme to work, it is required that all commitments be binding, meaning that once an offer is made, it cannot be revoked. Consequently, we analyse binding contracts as the commitment mechanism that enables \textit{game manipulation} scenarios. The main focus of this study is to formulate the logic of this setting, expand its scope to encompass more intricate schemes, and analyse the behaviour of regret-minimizing agents in scenarios involving game manipulation.
\end{abstract}


\sloppy

\newpage
\section{Introduction}

Commitment devices are tools that influence decision-making by linking actions to rewards or punishments. In the realm of game theory, the introduction of commitment devices leads to a distortion of payoff matrices, which in turn might result in the modification of the game's equilibrium \cite{Schelling1956,Sobel81,Vickers85}. For commitment devices to be effective, the offer made to modify a payoff matrix must 
% be binding. This means that once an offer is made, it cannot be revoked and it should be executed once the established conditions are met.
be credible. That is, the commitment to deliver punishments or rewards is either strategically self enforced \cite{Aumann1974}, or it is \emph{binding} --- meaning that once an offer is made, it cannot be revoked and it must be executed once the established conditions are met \cite{bono_wolpert_game_mining}.  

In this paper, we explore the logic of binding commitments in game theory. Thus, in the sequel, all offers made by the original players or additional actors, which we call \textit{game manipulators}, are binding. In contrast to previous literature, discussed methodically in \cref{sec:related_literature}, we concentrate on analysing the strategic implications of such commitment schemes.

For this, we assume that offers and actions are common knowledge to all participants. From a classical game theory perspective, it is also assumed that all players are rational. This implies that each player is driven by their individual self-interest, leading them to seek to maximise their own payoffs. Consequently, players select actions that align with the Nash equilibrium in an attempt to unilaterally maximise their payoffs.\footnote{Our analysis focuses on games with a single equilibrium. There are, however, cases where multiple equilibria arise in the manipulated game, in which case rationality does not resolve the equilibrium selection problem. In these cases, we study the convergence of no-regret dynamics via simulations.}

In such a framework, an actor, referred to as the \textit{game manipulator}, could enter and obtain a fee. The game manipulator simply needs to present the following binding offer to the players: pay me a fee, or I will distort the payoff matrices by benefiting or punishing a certain action. In other words, the manipulator obtains a profit in exchange for not distorting the game structure, effectively exploiting the game.

In this study, we systematically analyse the game-theoretical logic of obtaining fees from players by offering binding contracts. This provides fundamental insight into the potential actions within the realm of blockchain, where intricate systems have been established to guarantee the enforceability of contracts. Moreover, in such blockchain environments binding contracts, like smart contracts, have become the backbone of numerous decentralised activities.

%Such contracts, occurring on a blockchain, are referred to as smart contracts. They are designed to establish the terms of an agreement and automatically execute them once the specified conditions are met.
%A main advantage of smart contracts over traditional contracts is the reduced ambiguity coming from humans trying to interpret or negotiate contracts. In addition, it removes the need for trusted intermediaries to validate and enforce the contract once the conditions of the agreement are met.\\

We analyse diverse \textit{game manipulation} scenarios, including cases where one player makes an offer to its co-player in \cref{sec:player_to_player_offer}, cases where an external manipulator makes the offer in \cref{sec:First_miner_offer}, scenarios where the co-player responds with a counteroffer in \cref{sec:Player_counteroffer}, and finally, in \cref{sec:Second_miner_offer}, scenarios where a second external manipulator makes an offer to the first one according to the same principle.

In each scenario, we complement the simple theoretical analysis with a learning theory approach \cite{Fudenberg1998book,Hart2000} to investigate how agents learn to play games altered by binding contracts. Particularly, we use the multiplicative weights update algorithm \cite{Arora2012} and the replicator dynamics \cite{Hofbauer1998book} to explore how regret-minimizing agents \cite{Blum2007} behave under game manipulation. The purpose of this is to understand how agents, presumed to possess hindsight rationality \cite{Morrill2021}, act in such scenarios. This assumption is relevant to machine learning-driven agents \cite{Blum2005} and can serve as an initial model for human behaviour.

Overall, we observe that depending on the game structure, the dynamics may or may not converge to the predicted subgame perfect equilibria (SPE). This affects the overall contract incentives to the extent that a failure to reach the SPE suggests that external manipulators would have been better off refraining from offering the contracts initially. Thus, to effectively propose and profit from binding contracts, it is necessary to carefully examine the resulting game dynamics along with its strategic implications.

Morever, the results show that even though regret-minimizing agents effectively avoid dominated strategies, it is not guaranteed for them to converge to subgame perfect equilibria. Therefore, such refinement of the Nash equilibrium, is not necessarily in agreement with the evolutionary stable strategy of the replicator dynamics \cite{Selten1965,Smith1982book}. This implies that backward induction fails once again to predict evolutionary stable strategies \cite{Ramirez2023}.

\section{How to manipulate a game?}
\label{sec:how_to_mine}

A \textit{game manipulation} scenario requires the following conditions:
\begin{itemize}
\item Players have the potential to engage with one another through a game, called the original game or ``base game."
\item An actor, known as the ``game manipulator'', has the ability to request a fee for allowing such interaction to happen freely (without constraints on the players' actions), if the fee is paid.
\item Otherwise, the manipulator has the power to distort the payoff matrices by providing an incentive to a player for choosing a specific action.
\item Once an offer is made by the game manipulator, it cannot be revoked and it should be followed through. To be more precise, the offers are binding commitments.
\end{itemize}

% We focus our attention on cases where the threat that the game manipulator makes is to pay the players additional rewards for some of the game outcomes. In such cases, players cannot credibly decline these positive transfers.  
For simplicity, we study the case where a binding offer is proposed to only one player in a $2\times 2$ asymmetric game. (For a general formulation, we refer the reader to \cref{sec:appendix_generalisation}). 
As a result, game manipulation gives rise to a sequential game with the following steps:
\begin{enumerate}
\item An offer, which is a binding commitment, is proposed to player $i$.
\item Player $i$ must decide to Accept or Decline to pay the fee. If player $i$ Declines, then the player's payoff matrix is distorted as stated by the game manipulator.
\item Player $i$ plays the chosen game (given Accept or Decline) with its co-player.
\end{enumerate}

Essentially, player $i$ is faced with the alternative to either pay a fee, or have the game modified in such a way that the Nash equilibrium is shifted.

If the game is distorted to player $i$'s disadvantage, then player $i$ should pay the fee when the fee is smaller than that disadvantage. In such case, the manipulator is able to obtain ``free money" by only posing the threat to distort the payoff matrix by paying the players additional rewards. More precisely, the subgame perfect equilibrium in a game manipulation setting encourages player $i$ to Accept to pay the fee.

\subsection{Offer of a binding contract}
\label{sec:offer}
In this analysis, the offers are formulated according to the following format: \textit{Either you pay a fee, or if you decline, I distort the game.}

For example, the offer could be stated as:
\textit{Either you give me $K_1$ (euros), or if you decline, I give you $K_2$ whenever you play X.}

In this formulation, $X$ refers to an element of the offeree's action set. For example, in a $2 \times 2$ game, if the offer is proposed to the row player, $X$ is either Top or Bottom. Likewise, if the offer is proposed to the column player, $X$ is either Left or Right. It should be noted that the offeror can be either a player or an external actor.

\subsubsection{Importance of binding commitments}

It is important to note that the offers must be proposed as binding contracts, that is, there must be a mechanism in place to ensure that the offeror hands over $K_2$ when the request for $K_1$ is declined.

In traditional legal scenarios, the offeror could deposit $K_2$ with a notariser, who would return it to the offeror if the offeree's Accepts, but otherwise give it to offeree. However, in blockchain environments, such commitments can be made through smart contracts without the need of a third-party. As a result, game manipulation scenarios could proliferate more easily in blockchain platforms.

% \vspace{5pt}
% \noindent
All the offers analysed in this study use the following ``base game".

\section{The base game}
\label{sec:base_game}

Let us start with the following game

\begin{newnumbering}
\begin{center}
\begin{tabular}{c|c|c|}
&L&R \\
\hline
T& 4,14&9,13\\
\hline
B& 5,6& \textcolor{violet}{10,10} \\
\hline
\end{tabular}
\vspace{-1cm}
\bel{game1}
\qe
\vspace{.4cm}
\end{center}
\end{newnumbering}

The players are called Row (Row, he) and Column (Col, she). 
The unique Nash is $(Bottom,Right)$ with the payoff of $(10,10)$, which is Pareto-efficient. Furthermore, Bottom strictly dominates Top for the row player. Right is an iterated dominant strategy for the column player. Thus, the game is a one-sided dominant game.         

\subsection{Motivation}
For simplicity and tractability, we explore a $2\times2$ asymmetric game, where the binding offers are proposed to only one player. It should be noted that the game must be asymmetric such that the threat to distort the game is done by modifying only player $i$'s payoff matrix, as described in \cref{sec:offer}. This implies that the game manipulator has agency exclusively on player $i$ and not its co-player. Possible modifications of this formulation are discussed in section \cref{sec:appendix_generalisation} and \cref{sec:further_work}.

In particular, we propose game \rf{game1} as a base game because it is an asymmetric $2\times2$ that allows a tractable analysis of the multiple game manipulation scenarios presented in further sections.
Game \rf{game1} is also suitable to describe the buyer-supplier strategic interaction around demerit goods. This type of good is characterised by having negative effects on its consumers, i.e. demerit goods are considered to be unhealthy, demeaning or to have socially undesirable externalities, like pollution \cite{Robbins1932book, Boadway1984book}. As a result, buyers, and society, would be better off if demerit goods are produced and consumed in low quantities, or not at all.
To the contrary, suppliers/sellers are better off when they produce large, unrestricted, amounts of the demerit good in order to make sure they always meet the existent demand.

In particular, we assume that the buyers behave according to the supply and demand microeconomic model, such that the demand aims to match the quantity supplied. Precisely, economic equilibrium is reached once the demand coordinates with the supply. Following the interpretation of game \rf{game1}, the supply is better off by producing large amounts of the good to coordinate with the demand at high production levels. On the other hand, the demand prefers to coordinate with the supply at low levels of production, due to the negative effects of the transacted good.

Consequently, game \rf{game1} represents the buyer-supplier interaction for demerit goods, where Row is the supply and Column the demand, as follows.

\begin{newnumbering}
\begin{center}
\begin{tabular}{c|c|c|}
&Low consumption& High consumption \\
\hline
Low production& 4,14&9,13\\
\hline
High production& 5,6& \textcolor{violet}{10,10} \\
\hline
\end{tabular}
\vspace{-1cm}
\bel{game_example}
\qe
\vspace{.4cm}
\end{center}
\end{newnumbering}

The Nash equilibrium is (High production, high consumption).
In particular, High production is a strictly dominant strategy for the supply. On the other hand, the demand needs to best reply with high consumption, although it would strictly prefer Low production yielding always higher payoffs. 

Given that the Nash equilibrium is a socially undesirable equilibrium for the consumers, it is possible to model government intervention through \textit{game manipulation}. Similarly, a game manipulation scenario can be used to model extortive behaviour by dominant firms controlling the market, over smaller firms that have the potential to engage with their existing costumers.

\paragraph{Government intervention:} For a demerit good market, the government (external agent) proposes to the supplier (row player): \textit{Either you pay me $K_1$, or if you decline, I give you $K_2$ whenever you play Low production.}
This translates into the negotiation of governments with lobbyist from firms selling demerit goods. The government proposes the supplier to either Accept to pay a tax (\textit{Either you pay me $K_1$}), or to get a subsidy with the goal of reducing the production of the demerit good (\textit{or if you decline, I give you $K_2$ whenever you play Low production.}).

For example, the government can propose to the car industry selling vehicles powered by fossil fuels ``Either you pay me a tax, or if you decline I give you a subsidy to produce electric vehicles, as an incentive to decrease the production of petrol vehicles." Therefore, the analysis of game manipulation can provide a better understanding on how to implement more efficient regulations on markets.

\paragraph{Extortion from a dominant firm:} In a demerit good market, the dominant firm (external agent) proposes to a smaller firm (row player): \textit{Either you pay me $K_1$, or if you decline, I give you $K_2$ whenever you play Low production.}
This translates into an extortive threat from a dominant firm against a smaller one. The dominant firm pressures the smaller one to Accept to pay an extortive fee (\textit{Either you pay me $K_1$}), or the dominant firm will distort the market such that a Low production of the smaller firm is benefited (\textit{or if you decline, I give you $K_2$ whenever you play Low production.}).

In other words, the only way for the smaller firm to prevent that the dominant firm distorts the base game is to pay the extortion fee. This example serves to illustrate the wide variety of real-life situations where game manipulation acting as an extortion mechanism could come into place.

Furthermore, in the growing economy over blockchain platforms, any platform participant has the power to create binding contracts and may potentially intervene in a strategic interaction between other participants if the incentives to do so exist. However, unlike government interventions, game manipulation coming from extortionists is a concerning activity that is hard to regulate, especially in blockchain platforms.

Consequently, to provide a better understanding of the strategic implications of binding contracts, in the upcoming sections, we analyse various ways in which they can be introduced in games.

\vspace{5pt}
\noindent
We study the logic and dynamics of the following scenarios:

\begin{itemize}
  \item \cref{sec:player_to_player_offer}: The column player makes an offer to the row player
  \item \cref{sec:First_miner_offer}: The first external manipulator enters
  \item \cref{sec:Player_counteroffer}: Row player's counter-offer
  \item \cref{sec:Second_miner_offer}: The second external manipulator enters
\end{itemize}

\noindent
All offers are proposed taking the base game as the reference point. Regarding dynamics analysis, we employ the multiplicative weights and replicator dynamics to examine how regret-minimizing agents based on reinforcement learning behave in each scenario.

\section{Methods}
\label{sec:Methods}

We employ online algorithms to explore how agents learn to determine the best action in a game. In particular, we simulate the scenario in which agents cannot directly identify the best action from historical data alone; they must actively engage in the game to learn which action is optimal. For this, agents repeatedly play the game, aiming to reduce their cumulative losses, or more precisely, to minimize their regret \cite{Blum2007}.
In particular, we employ the most-studied no-regret algorithm, known as the multiplicative weights update algorithm (see \cite{Arora2012} and references therein), and its continuum limit \cite{Kleinberg2009}, the replicator dynamic \cite{Hofbauer1998book}.

\subsection{Multiplicative weights update algorithm}
\label{sec:Methods_MWUA}

The multiplicative weights algorithm (MWUA) is a decision making method based on an iterative rule that multiplicatively updates the weights assigned to each player's possible actions \cite{Littlestone1994, Fudenberg1995}. Initially, all actions have equivalent weights. After each time step, each action's weight is updated according to its payoff, thus the actions with higher payoffs have an increased probability of being chosen in the next round. More precisely, the players' learning process is modelled as a multiplicative weight update, where the action's distribution is skewed towards the successful actions providing a higher payoff \cite{Grigoriadis1995, Arora2012}. As a result, the main objective of the multiplicative weights algorithm is to reduce its cumulative losses, or more specifically, to minimize the regret for agents playing a repeated game \cite{Bailey2018}.
As stated in \cite{Arora2012}, the formulation of the algorithm is the following:

\vspace{5pt}
\noindent
Initialisation: fix $\eta \leq \frac{1}{2}$. Associate a weight $w_{i}^{(1)}:=1$ to each action $i$, for each player.

\noindent
For $t=1,2,...,T$:
 \begin{enumerate} 
  \item Choose action $i$ with probability proportional to its weight $w_{i}^{(t)}$, i.e. use the distribution $\textbf{p}^{(t)} = (\frac{w_1^{(t)}}{\Phi^{(t)}},..., \frac{w_n^{(t)}}{\Phi^{(t)}})$, where $\Phi^{(t)} = \sum_i w_i^{(t)}$, to choose an action on each round.
  \item Observe the payoffs of the actions $\textbf{m}^{(t)}$.
  \item Update the weights for every action as $w_i^{(t+1)}=w_i^{(t)}(1+\eta m_i^{(t)})$.
\end{enumerate}

\subsection{Replicator dynamics}
\label{sec:Methods_RepDyn}

The replicator dynamic is 
% the main no-regret dynamic in algorithmic game theory and 
a foundational model for dynamics in evolutionary game theory \cite{Hofbauer1998book, Hofbauer2009, Sandholm2010book}. In particular, it can be used as a model to analyse how players learn to play an asymmetric game.
In this dynamic, at each time step, the row player chooses one of the $n$ possible actions with frequency $\textbf{x} = (x_1, ..., x_n)$. Likewise, Column choose one of the $m$ possible actions with frequency $\textbf{y} = (y_1, ..., y_m)$. In our setting, Row has the same amount of possible actions as Column, $n=m$ (see base game \rf{game1}). Over time successful actions are positively reinforced, which allows for players to eventually converge to an optimum equilibrium.
Thus, the replicator dynamic is represented by the following set of coupled differential equations. 

\begin{equation}
    \dot{x_i} = x_i[(Ay)_i - x^TAy] \hspace{3mm} i=1,..,n
\label{rep_eq1}
\vspace{5pt}
\end{equation}
\begin{equation}
    \dot{y_j} = y_j[(Bx)_j - y^TBx] \hspace{3mm} j=1,..,m.
\label{rep_eq2}
\end{equation}

% \noindent 
In this equation, $A$ and $B$ are the payoff matrices of the row and column player, respectively. It should be noted that to obtain $B$, it is required to take the transpose of the game for the column entries, such that the definition of symmetric games holds: $A=B^T$.
For equation \ref{rep_eq1}, $(Ay)_i$ is Row's average performance of strategy $i$ over Column's possible actions. On the other hand, $x^TAy$ is the averaged performance over all of Row's strategies. Thus, the replicator equation describes how the frequency of strategy $i$ changes over time according to its average performance relative to the mean. Equation \ref{rep_eq2} has an equivalent interpretation for the column player \cite{Borgers1997, Hofbauer1998book,Sato2002}.

Dynamical systems such as the replicator dynamic arise as the continuum limit of discrete no-regret algorithms. More precisely, it has been shown that a multiplicative update process can be approximated to a continuous-time process displaying replicator dynamics \cite{Kleinberg2009, Kleinberg2011, Bailey2018}.

\section{Column makes an offer to Row}
\label{sec:player_to_player_offer}

To introduce the underlying logic of the binding commitments, we first allow for Col to make the following offer to Row: \emph{Either you give me 3 (Euros), or if you decline, I give you 2 whenever you play $T$.}

\noindent
If the row player Accepts, the payoff is as follows.

\begin{newnumbering}
\begin{center}
\begin{tabular}{c|c|c|}
  &L&R\\
\hline
 T& 1,17&6,16 \\
\hline
 B& 2,9&\textcolor{violet}{7,13}\\
\hline
\end{tabular}
\vspace{-1cm}
\bel{game2}
\qe
\vspace{.4cm}
\end{center}
\end{newnumbering}

The unique Nash equilibrium in this new game does not change compared to the base game \rf{game1}, as it remains at $(B,R)$, given that Bottom continues to be a strictly dominant strategy over Top. The main difference with respect to the base game is that now Col is able to extract a profit from Row, which in this case is equivalent to $3$.

\noindent
If the row player Declines, the payoffs are
\begin{newnumbering}
\begin{center}
\begin{tabular}{c|c|c|}
  &L&R\\
\hline
 T& \textcolor{violet}{6,12}&11,11 \\
\hline
 B& 5,6& 10,10 \\
\hline
\end{tabular}
\vspace{-1cm}
\bel{game3}
\qe
\vspace{.4cm}
\end{center}
\end{newnumbering}

Now the unique Nash in this new game is $(T,L)$, resulting in $(6,12)$. The Nash equilibrium changes because Top now becomes the strictly dominant strategy for the row player.

In particular, if the offer is Declined, Row's payoff matrix becomes that of a temptation game, that is, Row is tempted to play $T$ and obtain the fee offered by Col, but the resulting equilibrium leaves him worse off than if he had accepted the offer. Therefore, it is better for Row to Accept the offer and allow for Col to extract the profit. Likewise Col is better off if the Accept Nash equilibrium is reached. Most importantly, the column player is better off in either scenario, Accept or Decline, compared to the base game, which is the main motivation for Col to make the offer to Row.

Overall, for the full game, the subgame perfect equilibrium is the unique Nash equilibrium of the subgame \rf{game2}, corresponding to $(B,R)$ once Row Accepts given the NE in the subgame \rf{game3} with $(T,L)$ when Declining the offer. 

\subsection{Formulation conditions}
The binding offer to the row player can be generalised as: \emph{Either you give me $c_1$ (Euros), or if you decline, I give you $c_2$ whenever you play $T$.}
Therefore, an abstract representation of the offer can be formulated using the payoff matrix of the base game \rf{game1}

\begin{center}

\begin{tabular}{c|c|c|}
  &L&R \\
\hline
 T& $a_{11},b_{11}$& $a_{12},b_{12}$ \\
\hline
 B& $a_{21},b_{21}$&$a_{22},b_{22}$ \\
\hline
\end{tabular}
\vspace{-1cm}
\bel{game4}
\qe
%\vspace{.4cm}
\end{center}

% \begin{center}
\noindent
the Accept fee $c_1$ and the Decline fee $c_2$.  Particularly, we find that the crucial conditions are
\ba
\label{game5}
a_{12}+c_2>a_{22}>a_{12}\\
\label{game6}
a_{22}-c_1>a_{11} +c_2\\
\label{game7}
b_{11}>b_{12}\\
\label{game8}
a_{11}+c_2\ge a_{21}.
\ea
% \end{center}

\subsection{Reduced-Normal form game}
The sequential game composed of \rf{game2} and \rf{game3} can be written in its reduced-normal form, where Row has four possible actions (Accept and Top, Accept and Bottom, Decline and Top, Decline and Bottom). Similarly, Column has four possible actions but that depend on Row's initial decision between (Accept, Decline). For example, the action $L,R$ means that Column chooses Left if Row Accepts and Right if Row Declines.

\begin{newnumbering}
\begin{center}
\begin{tabular}{c|c|c|c|c|}
  &L,L&L,R&R,L&R,R\\
\hline
 AT& 1,17&1,17&6,16&6,16 \\
\hline
 AB& 2,9&2,9&\textcolor{violet}{7,13}&7,13 \\
\hline
 DT& \textcolor{violet}{6,12}&11,11&6,12&11,11 \\
\hline
 DB& 5,6&10,10&5,6&10,10 \\
\hline
\end{tabular}
\vspace{-1cm}
\bel{normalform1}
\qe
\vspace{.4cm}
\end{center}
\end{newnumbering}

The pure Nash equilibria are $(DT, \hspace{2mm} L,L)$ and $(AB,\hspace{2mm} R,L)$, which coincide with the equilibria of the individual games. There is also a mixed Nash equilibrium which we do not consider in our study. For this representation, it can also be noted that $AT$ and $DB$ are dominated strategies for Row.

\subsection{Results}
\label{sec:results_offer1}

\vspace{2mm}
\begin{figure}[h!]
\vspace{-16pt}
    \centering
    \includegraphics[width=1.0\textwidth]{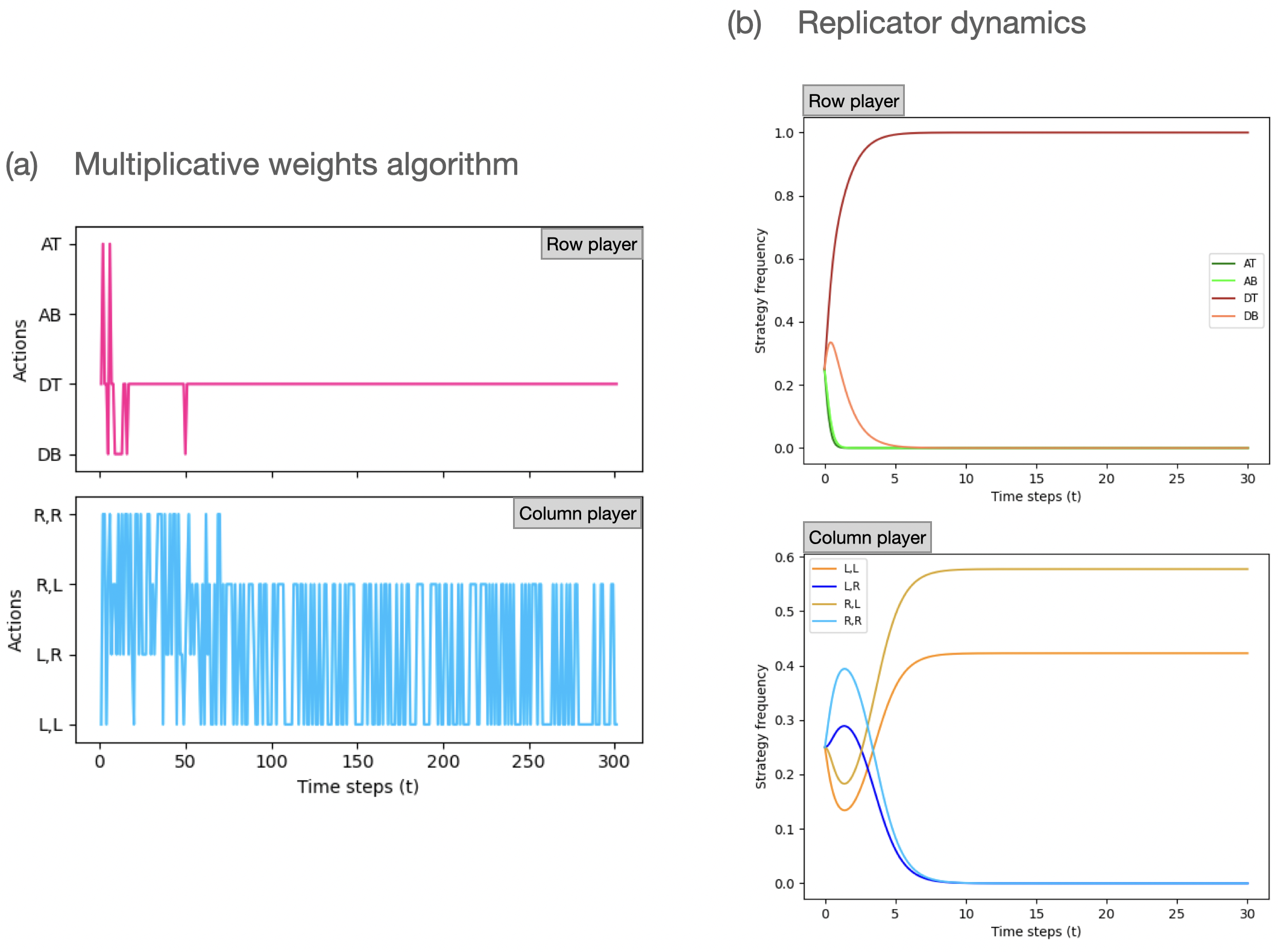}
    \caption{\textbf{(a)} A simulation of multiplicative weights dynamics ($\eta = 0.5$ for both players) and \textbf{(b)} numerical solution to the  replicator dynamics equations for \cref{sec:player_to_player_offer}, where Column makes an offer to Row. For both methods, the row player converges to $DT$, which means to Decline the offer and then playing Top. The column player converges to $L,L$ and $R,L$, corresponding to Col playing Left once Row has declined the offer. Thus, the system converges to the Nash equilibrium of the Decline game \rf{game3}. Interestingly, this equilibrium does not coincide with the subgame perfect equilibrium of the full game.
    }
    \label{fig:Results1}
\end{figure}

Figure \ref{fig:Results1} shows the results for the scenario where Column makes an offer to Row, as considered in the normal form game \rf{game9}. It can be seen that the multiplicative weights algorithm converges to the same equilibrium as the replicator dynamics, namely $(T,L)$, the Nash equilibrium of the Decline game. It should be noted that the MWUA plot shows the actions chosen by each player for every time step, while the replicator dynamics displays how the action frequency evolve in time. In general, both methods identify $DT$ as the best possible action for Row taking into account Column's available actions. It is surprising that the replicator dynamics and the MWUA equilibrium are consistently not in agreement with the subgame perfect equilibrium of the full game.

Interestingly, this suggests that Row prefers to Decline Column's offer, such that he is able to receive the $c_2=2$ euro offered to him. In this way Row exploits Column back by extracting the decline fee ($c_2$) from her. The reason for such behaviour is that $DT$ 
performs better on average with respect to all possible strategies of the column player. Unfortunately, if the row player Declines the drawback is that the game structure leaves him worse off compared to the Accept equilibrium.

\section{The first external manipulator enters}
\label{sec:First_miner_offer}

The observation that players can change a game to their advantage by making binding commitments dates back to \cite{Schelling1956}. Thus, we now introduce a novel scenario where an external ``game manipulator" ($M_1$), makes an offer, in the form of a binding contract, to extort a fee from players. More precisely, the game manipulator makes the offer to Row: \emph{Either you give me 3 (Euros), or if you decline, I give you 2 whenever you play $T$.}

It should be noted that the main differences with respect to the first scenario is that an external game manipulator is making the offer, instead of a player. Therefore, payoffs injected or retrieved by the manipulator are not transfers between the two players anymore. The equilibrium strategies for Row and Column remain, however, the same.  

\noindent
If the row player Accepts, the payoff matrix is as follows

\begin{newnumbering}
\begin{center}
\begin{tabular}{c|c|c|}
  &L&R\\
\hline
 T& 1,14,\{3\}&6,13,\{3\} \\
\hline
 B& 2,6,\{3\}& \textcolor{violet}{7,10,\{3\}} \\
\hline
\end{tabular}
\vspace{-1cm}
\bel{game9}
\qe
\vspace{.4cm}
\end{center}
\end{newnumbering}
The third entry in a cell specifies the payoffs to the external manipulator $M_1$. It should be noted that after the offer is proposed, it cannot be revoked. Thus, the payoff of $M_1$ is solely an outcome of the players' (Row and Col) decisions, such that the Nash equilibrium in this subgame is $(B,R)$ with payoffs $(7,10,\{3\})$. For Row, Bottom is a strictly dominant strategy while Column has an iterated dominant strategy. If Row Accepts, the game manipulator extracts a profit of $c_1=3$.

If the row player Declines, the payoffs are

\begin{newnumbering}
\begin{center}
\begin{tabular}{c|c|c|}
  &L&R\\
\hline
 T& \textcolor{violet}{6,14,\{-2\}}&11,13,\{-2\} \\
\hline
 B& 5,6,\{0\}&10,10,\{0\} \\
\hline
\end{tabular}
\vspace{-1cm}
\bel{game10}
\qe
%\vspace{.4cm}
\end{center}
\end{newnumbering}

The Nash equilibrium in this subgame is $(T,L)$ with payoffs $(6,14,\{-2\})$. In this game, Top is a strictly dominant strategy. Left is an iterated dominant strategy. Therefore, if Row Declines, the game manipulator pays a fee of $c_2=2$ in this subgame.

The external game manipulator does not change the strategic situation for Row compared to the previous case in \cref{sec:player_to_player_offer}, where Col made the offer. Again, it is better for Row to Accept the offer, and pay $c_1=3$, to reach the equilibrium in the Accept-subgame, $(B,R)=(7,10,\{3\})$, instead of the equilibrium in the Decline-subgame $(DT,L)=(6,14,\{-2\})$. In other words, the subgame perfect equilibrium outcome corresponds to the Nash equilibrium outcome of game \rf{game9}, namely $(B,R)$ once Row Accepts.

It should be noted that for the logic of the external manipulator scenario, it is not necessary for the fee the manipulator demands ($c_1=3$) to be greater than the fee he threatens to pay ($c_2=2$). As long as it is better for Row to Accept the offer, any positive fee $c_2$ would be sufficient. $c_2$ just cannot be too large, according to \rf{game9}, or else the subgame perfect equilibrium would be modified and Row would be better off Declining.

\subsection{Reduced-Normal form game}
The sequential game described by the games \rf{game9} and \rf{game10} can be described in its reduced-normal form as

\begin{newnumbering}
\begin{center}
\begin{tabular}{c|c|c|c|c|}
  &L,L&L,R&R,L&R,R\\
\hline
 AT& 1,14,\{3\}&1,14,\{3\}&6,13,\{3\}&6,13,\{3\} \\
\hline
 AB& 2,6,\{3\}&2,6,\{3\}&\textcolor{violet}{7,10,\{3\}}&7,10,\{3\} \\
\hline
 DT& \textcolor{violet}{6,14,\{-2\}}&11,13,\{-2\}&6,14,\{-2\}&11,13,\{-2\} \\
\hline
 DB& 5,6,\{0\}&10,10,\{0\}&5,6,\{0\}&10,10,\{0\} \\
\hline
\end{tabular}
\vspace{-1cm}
\bel{normalform2}
\qe
\vspace{.4cm}
\end{center}
\end{newnumbering}

The pure Nash equilibria are $(DT, \hspace{2mm} L,L)$ and $(AB,\hspace{2mm} R,L)$, which coincide with the equilibria of the individual subgames. There is also a mixed Nash equilibrium which we do not consider here. It can be noted that $AT$ and $DB$ are dominated strategies for Row.

\subsection{Results}
\label{sec:results_offer2}
\begin{figure}[h!]
\vspace{-21pt}
    \centering
    \includegraphics[width=1.0\textwidth]{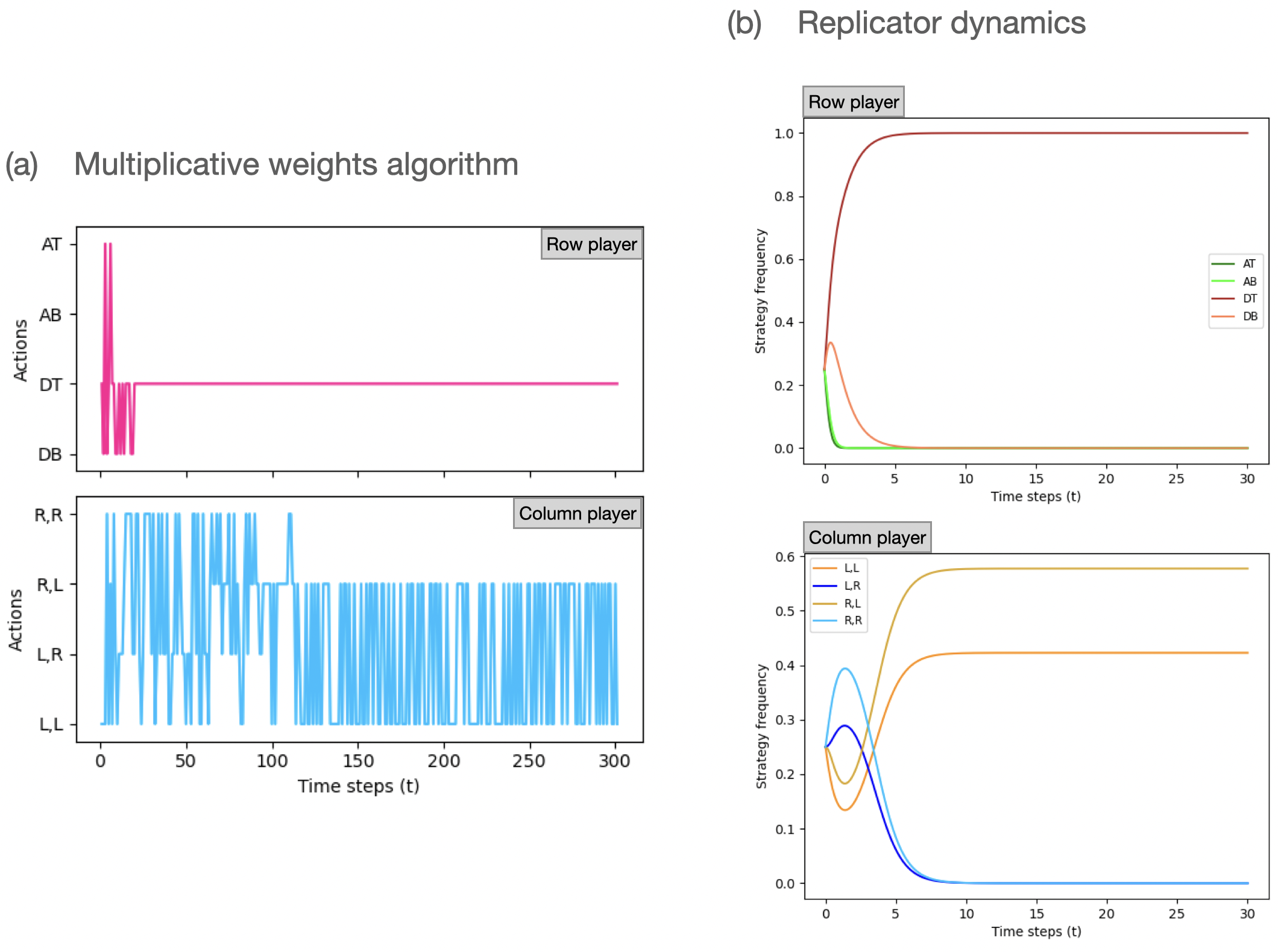}
    \caption{\textbf{(a)} A simulation of multiplicative weights dynamics ($\eta = 0.5$ for both players) and \textbf{(b)} numerical solution to the  replicator dynamics equations for \cref{sec:First_miner_offer}, where the first external manipulator enters. For both methods, the row player converges to $DT$, which means to Decline the offer and then playing Top. The column player converges to $L,L$ and $R,L$, corresponding to Col playing Left once Row has declined the offer. Thus, the system converges to the Nash equilibrium of the Decline game \rf{game10}. Interestingly, this equilibrium does not coincide with the subgame perfect equilibrium of the full game.}
    \label{fig:Results2}
    \vspace{-4pt}
\end{figure}

Figure \ref{fig:Results2} shows that the system converges to Row Declining the offer and then playing Top, and to Column playing Left. More precisely, the system converges to the Nash equilibrium of game \rf{game10}. The results are equivalent to the scenario where Column makes an offer to Row. The reason for this is that the general payoff structure is not changed when an external manipulator makes the offer instead of the column player. Therefore, once again we reach an equilibrium that does not coincide with the subgame perfect equilibrium of the full game. In this case, the results mean that Row takes advantage of the situation to extract a fee $c_2$ from the manipulator at the cost of having a lower payoff than the base game, and the Accept equilibrium. This decision indirectly benefits Col who obtains a better payoff for the Decline equilibrium in comparison to the base game and the Accept equilibrium.

 It is interesting to note that the results in figure \ref{fig:Results2} do not correspond to the subgame perfect equilibrium of the full game. The reason for this is the fact that on average $DT$ performs better against all of Column's possible actions. Therefore, the results indicate that despite the theoretical prediction of players reaching the subgame perfect equilibrium, which would enable the external manipulator to earn a profit of $c_1$, the manipulator would actually find it more advantageous to abstain from making an offer. This is because players tend to converge towards the Decline equilibrium, which incurs a cost of $c_2$ for the manipulator.

\section{Row's counter-offer}
\label{sec:Player_counteroffer}

In the scenario of the first external manipulator, Row could propose a counter-offer to Col: \emph{I shall decline the external manipulator's offer and then give you $d_2=2$ for you playing Right}. Therefore, if the row player Accepts, the payoff matrix is the same as game \rf{game9}, namely

\begin{newnumbering}
\begin{center}   
\begin{tabular}{c|c|c|}
  &L&R\\
\hline
 T& 1,14,\{3\}&6,13,\{3\} \\
\hline
 B& 2,6,\{3\}& \textcolor{violet}{7,10,\{3\}} \\
\hline
\end{tabular}
\vspace{-1cm}
\bel{game11}
\qe
\vspace{.4cm}
\end{center}
\end{newnumbering}

\noindent
The unique Nash equilibrium in this subgame is again $(B,R)=(7,10,\{3\})$, given that Bottom is a strictly dominant strategy and Right is iterated dominant. In this case, the external game manipulator extracts a profit of $c_1=3$.

\noindent
If the row player Declines, Row's counter-offer changes the payoff matrix as follows

\begin{newnumbering}
\begin{center}
\begin{tabular}{c|c|c|}
  &L&R\\
\hline
 T& 6,14,\{-2\}& \textcolor{violet}{9,15,\{-2\}}\\
\hline
 B& 5,6,\{0\}&8,12,\{0\} \\
\hline
\end{tabular}
\vspace{-1cm}
\bel{game12}
\qe
\vspace{.4cm}
\end{center}
\end{newnumbering}

\noindent
Therefore, in subgame \rf{game13} $(T,R) = (9,15,\{-2\})$ is the new Nash equilibrium, where Top and Right are strictly dominant strategies. In this equilibrium, Row and Col obtain the best possible outcome. Even though Row is worse off than in the base game \rf{game1}, it is better off than in the Accept game and the previous Decline equilibrium. For Column, it is beneficial to accept the offer because she is also better off than in those two cases. In other words, Row has essentially transferred the fee obtained from the external manipulator to Col. But if that is possible, the manipulator should not attempt to pose a threat in the first place, as he risks losing $c_2=2$ by his attempt. Of course, this also depends on certain additional inequalities, with $d_2$ now denoting the payoff increase offered by Row to Col for playing Right.
\ba
\label{game13}
b_{12}<b_{11} < b_{12}+d_2\\
\label{game14}
a_{11}+c_2 < a_{12}-d_2
\ea
Thus, Row's counter-offer can prevent the game manipulator from extracting a profit. 

\subsection{Reduced-Normal form game}
The sequential game described by the games \rf{game11} and \rf{game12} can be represented in its reduced-normal form as

\begin{newnumbering}
\begin{center}
\begin{tabular}{c|c|c|c|c|}
  &L,L&L,R&R,L&R,R\\
\hline
 AT& 1,14,\{3\}&1,14,\{3\}&6,13,\{3\}&6,13,\{3\} \\
\hline
 AB& 2,6,\{3\}&2,6,\{3\}&\textcolor{violet}{7,10,\{3\}}&7,10,\{3\} \\
\hline
 DT& 6,14,\{-2\}&\textcolor{violet}{9,15,\{-2\}}&6,14,\{-2\}&\textcolor{violet}{9,15,\{-2\}} \\
\hline
 DB& 5,6,\{0\}&8,12,\{0\}&5,6,\{0\}&8,12,\{0\} \\
\hline
\end{tabular}
\vspace{-1cm}
\bel{normalform2}
\qe
\vspace{.4cm}
\end{center}
\end{newnumbering}

The pure Nash equilibria are $(DT, \hspace{2mm} L,R)$, $(DT, \hspace{2mm} R,R)$ and $(AB,\hspace{2mm} R,L)$, which coincide with the equilibria of the individual subgames games. It can be noted that $AT$ and $DB$ are dominated strategies for Row.

\subsection{Results}
\label{sec:results_offer3}

\begin{figure}[h!]
    \centering
    \includegraphics[width=1.0\textwidth]{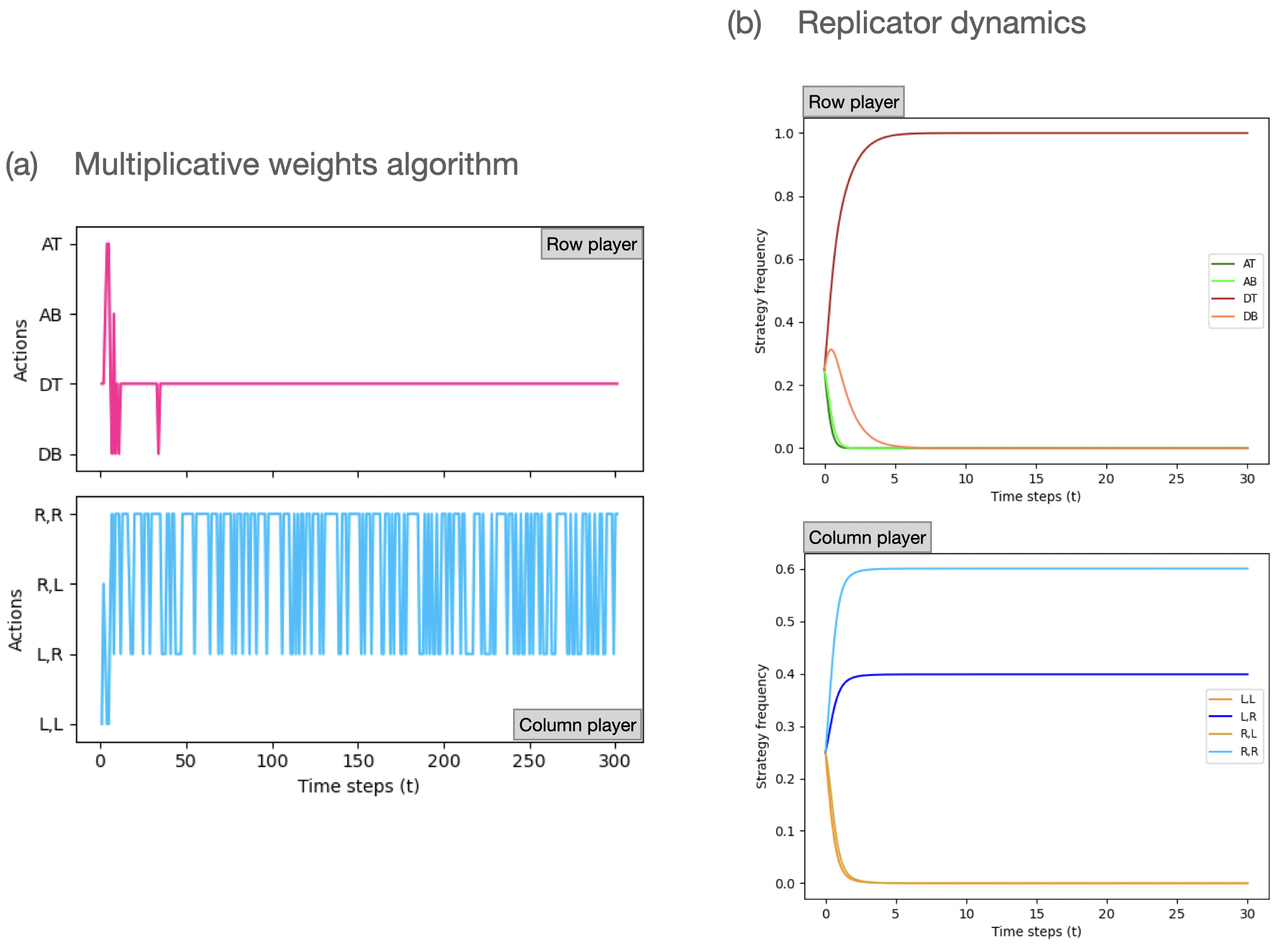}
    \caption{\textbf{(a)} A simulation of multiplicative weights dynamics ($\eta = 0.5$ for both players) and \textbf{(b)} numerical solution to the replicator dynamics equations for \cref{sec:Player_counteroffer}, which describes Row's counter-offer. For both methods, the row player converges to $DT$, which means to Decline the offer and then playing Top. The column player converges to $R,R$ and $L,R$, corresponding to Col playing Right once Row has declined the offer. Thus, the system converges to the Nash equilibrium of the Decline game \rf{game12}, which is in agreement with the subgame perfect equilibrium. Such equilibrium allows for both Row and Column to obtain the best possible payoff.}
    \label{fig:Results3}
    \vspace{-7pt}
\end{figure}

As shown in figure \ref{fig:Results3}, for the multiplicative weights algorithm and the replicator dynamics, the system converges to Row Declining and playing Top, and then for Column to accept Row's counter-offer to play Right. Therefore, the results are in agreement with the subgame perfect equilibrium. This means that Row's counter-offer effectively neutralises the external game manipulator to the extent that the manipulator would be better off by not posing a threat on the first place.

\section{The second external manipulator enters}
\label{sec:Second_miner_offer}

In this section, we introduce a second external manipulator into \cref{sec:First_miner_offer}, where the first external manipulator appears. Thus, the first manipulator's offer is the same as in section \cref{sec:First_miner_offer}.
The second manipulator $M_2$ (she) offers to the first manipulator $M_1$ (he): \emph{Either you pay me 2, or else if Row Declines your offer, I will offer Col an additional payoff of 2 for playing Right.}

The chronology of this scenario is as follows:
\begin{enumerate}
    \item Time $t_1$: Row and Column agree to play the original game, at time $t_6$.
    \item Time $t_2$: $M_1$ proposes the offer described in \cref{sec:First_miner_offer} to Row, specifying that they must make their decision precisely at time $t_5$.
    \item Time $t_3$: $M_2$ proposes the offer described in this section to $M_1$, specifying that they must make their decision precisely at time $t_4$.
    \item Time $t_4$: $M_1$ decides upon the offer made by $M_2$.
    \item Time $t_5$: Row decides upon the offer made by $M_1$.
    \item Time $t_6$: Row and Column play the game determined in previous steps.
\end{enumerate}

We focus on analysing the strategic implications of the offer from $M_2$ to $M_1$ (steps $t_3$ and $t_4$).

If $M_1$ accepts $M_2$'s offer, it implies that the Accept game is equivalent to game \rf{game9}, but $M_1$'s payoff is decreased by 2 amounting now to 1. Additionally, $M_2$'s payoff of 2 is added to the matrix, such that each entry follows the structure $($Row, Col, $\{M_1, M_2\})$.

\begin{newnumbering}
\begin{center}
\begin{tabular}{c|c|c|}
  &L&R\\
\hline
 T& 1,14,\{1,2\}&6,13,\{1,2\} \\
\hline
 B& 2,6,\{1,2\}& \textcolor{violet}{7,10,\{1,2\}} \\
\hline
\end{tabular}
\vspace{-1cm}
\bel{game13}
\qe
\vspace{.4cm}
\end{center}
\end{newnumbering}

\noindent
The Nash equilibrium in this subgame is $(B,R)$ with payoffs $(7,10,\{1,2\})$. For Row, Bottom is a strictly dominant strategy and for Col, Right is iterated dominant. The main difference in comparison to \cref{sec:First_miner_offer} is that now $M_1$ only extracts a profit equivalent to $1$, instead of $3$, and $M_2$ appears with a payoff of 2. 

\noindent
If $M_1$ declines $M_2$'s offer and Row Declines $M_1$'s offer, the new Decline subgame is as follows:

\begin{newnumbering}
\begin{center}
\begin{tabular}{c|c|c|}
  &L&R\\
\hline
 T& 6,14,\{-2,-2\}& \textcolor{violet}{11,15,\{-2,-2\}} \\
\hline
 B& 5,6,\{0,0\}&10,12,\{0,0\} \\
\hline
\end{tabular}
\vspace{-1cm}
\bel{game14}
\qe
\vspace{.4cm}
\end{center}
\end{newnumbering}

\noindent
This leads to the Nash equilibrium in this subgame $(T,R)=(11,15,\{-2,-2\})$ which is the best that Row and Col can hope for. In particular, Top and Right are strictly dominant strategies.

In general, the logic of the scenario is the following: $M_1$ makes an offer to Row. Either Row accepts $M_1$'s offer, or the Nash equilibrium is shifted to Row's disadvantage. It is thus better for Row to accept $M_1$'s offer. But before Row decides, $M_2$ enters and makes the offer to $M_1$. Either $M_1$ accepts $M_2$'s offer, or the game is changed in such a way that it now becomes advantageous for Row to decline $M_1$'s offer. Therefore, it is best for $M_1$ to accept $M_2$'s offer, and when Row then decides about $M_1$'s offer, it is also better to accept. 

In fact, the subgame perfect equilibrium states that Row would be better off by Declining $M_1$'s offer, in case $M_1$ declines $M_2$'s offer. Therefore, the threat posed by $M_2$ to distort the game, by offering an additional payoff to the column player, guarantees that $M_1$'s offer is Declined by Row. As a result, $M_1$ is forced to pay $M_2$ a fee to retain a profit from initially manipulating the Row player.

Interestingly, the results in \cref{sec:First_miner_offer} show that regret-minimizing players do not converge initially to Accept $M_1$'s offer, i.e. $M_1$'s profit is not guaranteed from the start. As a result, the current situation presents an interesting scenario, where a static analysis of the game contradicts the dynamical outcome obtained from online learning of regret-minimizing agents.

\subsection{Reduced-Normal form game}
The sequential game described by the games \rf{game13} and \rf{game14} can be represented in its reduced-normal form as

\begin{newnumbering}
\begin{center}
\begin{tabular}{c|c|c|c|c|}
  &L,L&L,R&R,L&R,R\\
\hline
 AT& 1,14,\{1,2\}&1,14,\{1,2\}&6,13,\{1,2\}&6,13,\{1,2\} \\
\hline
 AB& 2,6,\{1,2\}&2,6,\{1,2\}&\textcolor{violet}{7,10,\{1,2\}}&7,10,\{1,2\} \\
\hline
 DT& 6,14,\{-2,-2\}&\textcolor{violet}{11,15,\{-2,-2\}}&6,14,\{-2,-2\}&\textcolor{violet}{11,15,\{-2,-2\}} \\
\hline
 DB& 5,6,\{0,0\}&10,12,\{0,0\}&5,6,\{0,0\}&10,12,\{0,0\} \\
\hline
\end{tabular}
\vspace{-1cm}
\bel{normalform3}
\qe
\vspace{.4cm}
\end{center}
\end{newnumbering}

\noindent
Note that we do not specify the actions by $M_1$ and $M_2$ explicitly. In particular, the pure Nash equilibria are $(DT, \hspace{2mm} L,R)$, $(DT, \hspace{2mm} R,R)$ and $(AB,\hspace{2mm} R,L)$, which coincide with the equilibria of the individual subgames. It can be noted that $AT$ and $DB$ are dominated strategies for Row.

\subsection{Results}
\label{sec:results_offer4}

\begin{figure}[h!]
    \centering
    \includegraphics[width=1.0\textwidth]{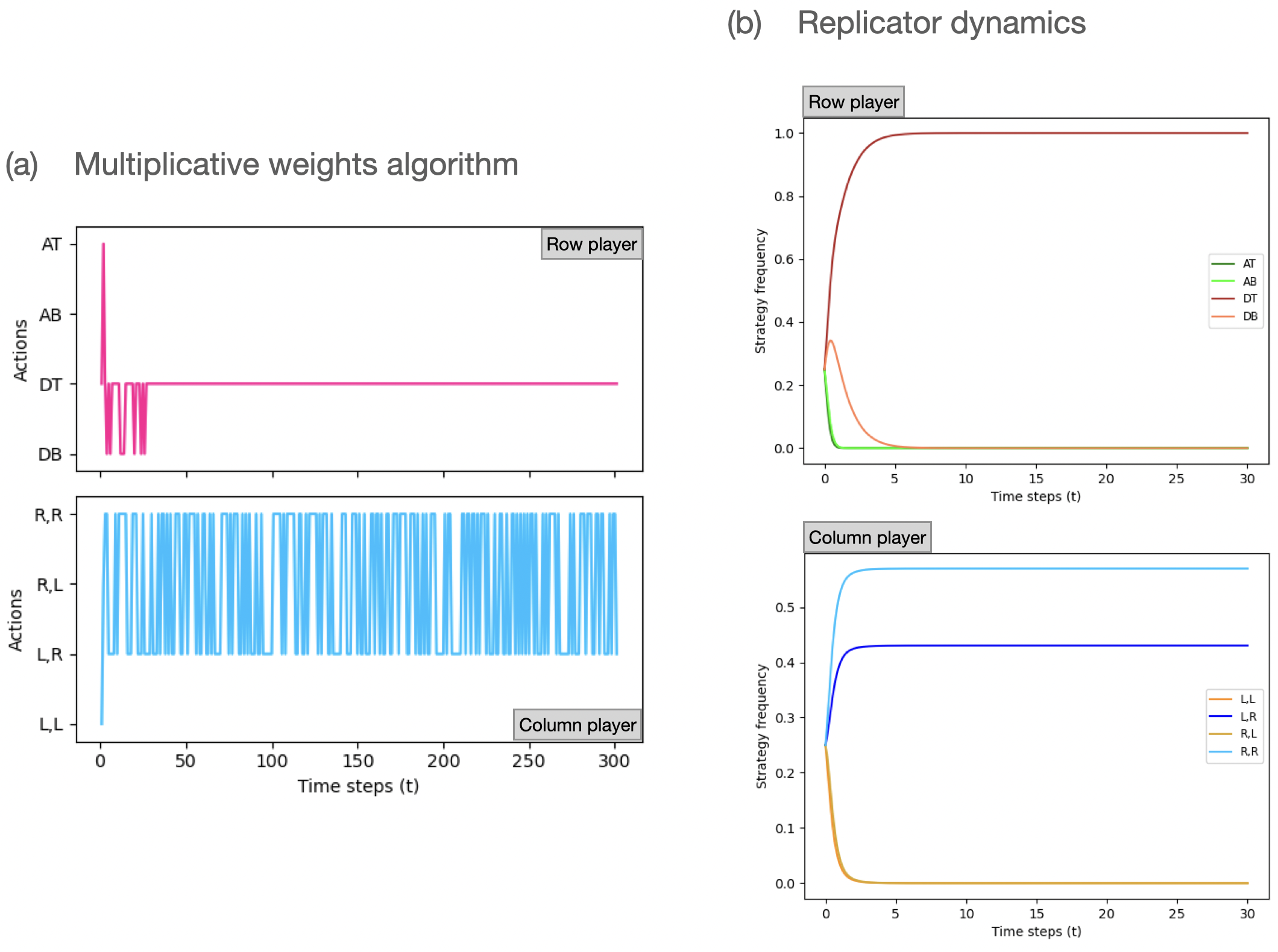}
    \caption{\textbf{(a)} A simulation of multiplicative weights dynamics ($\eta = 0.5$ for both players) and \textbf{(b)} numerical solution to the  replicator dynamics equations for \cref{sec:Second_miner_offer}, where the second external manipulator enters. For both methods, the row player converges to $DT$, which means to Decline the offer and then playing Top. The column player converges to $R,R$ and $L,R$, corresponding to Col playing Right once Row has declined the offer. Thus, the system converges to the Nash equilibrium of the Decline game \rf{game14}. If $M_2$ offer is common knowledge, this equilibrium is equivalent to the subgame perfect equilibrium. This means that $M_1$ is effectively forced to pay a fee to $M_2$ to be able to retain a profit from initially manipulating Row. However, this is a paradoxical result because $M_1$'s profit is not guarantee from the start when regret-minimizing agents are playing the game.}
    \label{fig:Results4}
\end{figure}

In figure \ref{fig:Results4}, it can be seen that the system converges to Row Declining the offer and playing Top, while Column plays Right. This equilibrium coincides with the Nash equilibrium of game \rf{game14}. Therefore, if $M_2$'s offer is common knowledge, the players converge to the subgame perfect equilibrium, where both Row and Col reach the best possible outcome. This means that $M_1$ is effectively forced to pay a fee to $M_2$ to be able to retain a profit from initially manipulating Row.

The current situation presents an interesting paradox when compared to \cref{sec:First_miner_offer}, because the results indicate that regret-minimizing players do not converge initially to Accept $M_1$'s offer. This means that $M_1$'s profit is not guarantee from the start. As a result, we find ourselves in an intriguing scenario where a static analysis of the game contradicts the dynamical outcome obtained from regret-minimizing agents.

\section{Conclusions}

In this paper, we explore commitment devices in game theory. More precisely, we study the strategic implications of binding contracts using a tractable $2 \times 2$ asymmetric game. Given that game manipulation through binding contracts can take different forms, from government regulations to extortion in blockchain platforms, we systematically analyse the logic of \textit{game manipulation} via diverse setups.

Particularly, we investigate how regret-minimizing agents interact under the influence of game manipulators. For this, we use the multiplicative weights update algorithm (MWUA) and the replicator dynamics. We found that both methods provide consistent results for all analysed scenarios. 
% The reason for this is that the replicator dynamic is the continuum limit of discrete no-regret procedures, as the MWUA.
In particular, the results show that the regret-minimizing agents avoid convergence to dominated strategies, as could be expected.

Furthermore, we observe that depending on how the binding offers distort the base game, regret-minimizing agents might not converge to the subgame perfect equilibrium. For instance, in both \cref{sec:player_to_player_offer}, where Column makes an offer to Row, and \cref{sec:First_miner_offer}, where the first external manipulator enters, players converge to the Decline equilibrium, $(T,L)$, while the subgame perfect equilibrium corresponds to the Accept equilibrium, namely $(B,R)$ once Row Accepts. This indicates that regret-minimizing agents do not necessarily behave as backward induction predicts. 

In addition, the results reveal that despite the distinction between \cref{sec:player_to_player_offer}, where Column proposes the offer Row, and section \cref{sec:First_miner_offer}, where an external manipulator ($M_1$) proposes the offer to Row, both scenarios are dynamically equivalent for MWUA and the replicator dynamics.
For both methods, the row player prefers to Decline the offer, such that he is able to extract the fee $c_2$ offered to him. The reason for such behaviour is that $DT$ performs better on average with respect to all possible strategies of the column player. Unfortunately, the drawback is that if Row Declines, the game structure leaves him worse off compared to the Accept equilibrium.

This suggests a contradicting situation for $M_1$, because classical game theory suggests that players would achieve the subgame perfect equilibrium and allow him to extract a profit equivalent to $c_1$. However, the equilibrium achieved by regret-minimizing agents indicate that refraining from offering a contract would actually be more beneficial for $M_1$. Thus, when offering binding contracts to regret-minimizing agents, or agents only rational at hindsight, it is essential to carefully examine the dynamic implications of the contracts, in order to effectively leverage them. Relying solely on an analysis rooted in classical game theory may not be adequate for this purpose.

On the other hand, in both \cref{sec:Player_counteroffer}, describing Row's counter-offer, and \cref{sec:Second_miner_offer}, where the second external manipulator enters, MWUA and the replicator dynamics converge to the subgame perfect equilibrium. The reason for this is that both offers alter the game structure in a similar manner, leading to two pure Nash equilibria, in the normal form game, corresponding to Declining the offer. Therefore, despite their differences, both offers have an equivalent impact on the game structure and its dynamical outcome.

In the case of \cref{sec:Player_counteroffer}, the results indicate that Row's counter-offer to Column effectively neutralises the external game manipulator $M_1$ intention to profit from posing a threat to distort the game. Particularly, the game manipulator $M_1$ would be better off by not proposing a binding contract on the first place under Row's counter-offer.

Regarding \cref{sec:Second_miner_offer}, the second external manipulator $M_2$ proposes the binding contract to $M_1$ with the intention of extracting a fee from him. To be more precise, when $M_1$ Declines the offer of $M_2$, Row would be better off by also Declining the offer of $M_1$. This would inflict a loss of $c_2$ to $M_1$, whereas if $M_1$ accepts the offer of $M_2$, he would retain a profit, because in that case, Row should Accept $M_1$'s offer according to classical game theory. Moreover, the results obtained from MWUA and the replicator dynamics confirm that $M_2$'s offer is a credible threat for $M_1$.

Interestingly, \cref{sec:Second_miner_offer}, where the second external manipulator enters, poses a paradox in the light of the results of \cref{sec:First_miner_offer}, where the first external manipulator enters. The reason is that regret-minimizing agents converge to decline $M_1$'s offer, thus $M_1$'s profit obtained from initially manipulating Row is not guaranteed from the start. Therefore, $M_2$'s intervention becomes unreasonable under such consideration.

In conclusion, a static analysis based solely on the concept of Nash equilibrium, an its refinement the subgame perfect equilibrium, is not sufficient to understand the behaviour of regret-minimizing agents in the framework of game manipulation through binding contracts. A dynamic exploration is therefore crucial to effectively leverage binding contracts as commitment devices in game theory.

\section{Extensions and open questions}
\label{sec:further_work}

Possible extensions of the presented framework include bargaining over the details of the contract. In that bargaining, the manipulator attempts to maximise the ``free money" that a player could give it. This topic is related to mechanism design applied to the offer of the binding contracts. In the paper by \cite{bono_wolpert_game_mining}, the authors derive on the maximal amount a game manipulator can obtain for certain game manipulation scenarios, as one varies over possible payoff matrices of the base game and/or over possible contracts offered by the game manipulator. The analysis in that paper also shows that the results of game manipulation on the resultant joint payoffs of the players in the base game may differ in general from the Coaseian outcome. 

In all of these cases, the presence of the game manipulator results in an entirely different strategic environment faced by the players in the base game. Therefore, the manipulator makes considerable profit from recognising a situation in which one party benefits from an output-contingent contract. But if the game manipulator manages to make a profit, the possibility that other game manipulators appear, seeking to pose their own contracts at lower prices, arises. This opens additional strategic implications for the player whose payoff matrix is distorted by such offers.

Another option is for the external game manipulator to offer contracts to both players.  As an example, it may be that the manipulator offers contracts to both players with the following properties. First, both players have a strictly dominant strategy to accept the offered contract. Second, when they both inevitably accept, the outcome is that they are both worse off, and the game manipulator profits considerably. In fact, the game manipulator may even promise to pay the players large sums of money if certain outcomes are reached, knowing well that when all of the players accept the offered contracts,such outcomes will never be obtained. (To do this, the game manipulator effectively creates a prisoner's dilemma among the players for his own benefit.)

It is also worth noting that in many game manipulation scenarios, ``mixed contracts" are needed to guarantee the existence of equilibria. With such contracts, players have uncertainty about their opponent's payoffs. However, unlike in Bayesian Nash equilibrium, for game manipulation scenarios this uncertainty is resolved before the game is played. Thus, in this setting, the game manipulator's only role is to make the players indifferent among its own contracts such that players randomise among them according to a probability distribution.

Furthermore, a rich mathematical structure is introduced if the players in the base game are able to offer contracts to the external game manipulator, who can pick only one of them, in which those players ``purchase" a move-contingent payoff after the game is played.

Finally, there are important timing issues in game manipulation scenarios, some of which are explored in \cite{bono_wolpert_game_mining}. When players sequentially sign contracts with game manipulators, there can be a significant first-mover advantage to the first-signing player. This provides the game manipulator with yet another opportunity for profit i.e. they can charge the players to move first. In addition, game manipulation has complicated effects on efficiency. The effects depend on the underlying game as well as the bargaining structure between players and the game manipulator. In general, those effects can either increase equilibrium efficiency or decrease it.

% \bigskip
% {\bf Funding:}\\

\subsection*{Acknowledgments} David Wolpert thanks the Santa Fe Institute for support. This project has received funding from the European Research Council (ERC) under the European Union's Horizon 2020 Research and Innovation Programme (grant agreement no. 740282).
Rosemarie Nagel acknowledges the Spanish Ministry of Education grant PRESP04022 - PID2021-125538NB-I00/AEI/10.13039/501100011033 and the support by the Barcelona School of Economics. 

\section{Appendix}
\subsection{Generalised game manipulation}
\label{sec:appendix_generalisation}
For the general formulation, originally proposed in \cite{bono_wolpert_game_mining} (there termed ``game mining''), we first introduce the corresponding notation. Let $T_1 \in \{Row, Col\}$ be the unit for information, or bit, specifying which of the two players, Row or Col, the game manipulator asks for a one-time payment in a single-stage game, before the start of the base game. Let $T_2 \in \{Row, Col\}$ be a bit specifying which of the two players the manipulator will make an output-contingent payment to, if the request for a one-time payment is refused. Let $W \in \{win, lose\}$ be the bit saying whether the player specified in $T_1$ has better or worse expected payoff due to the intervention of the game manipulator, compared to what their expected payoff in the base game would be if the manipulator had never intervened. So for example, in the standard game manipulation scenario considered in \cref{sec:First_miner_offer}, $T_1 = T_2 = Row$ and $W = lose$; the player being exploited by the manipulator is worse off due to the intervention of the manipulator.

Nonetheless, as pointed out by \cite{bono_wolpert_game_mining}, there are base games in which a game manipulator could benefit by intervening for either or both possibilities, $T_1 \ne T_2$ and $T_1 = T_2$. Moreover, there are games where $W = win$ and games in which $W = lose$, i.e., the player in the base game who is being exploited by the manipulator can be made either worse off or better off. Thus, the particular case presented in \cref{sec:First_miner_offer} is just one possible configuration in the myriad of all available game manipulation opportunities.

\subsection{Game manipulation compared to previous literature}
\label{sec:related_literature}

The ideas underlying the game manipulation concept are implicit in a large body of economic literature. As an illustration, in the model presented in \cite{Jackson05} (JW), every player specifies outcome-contingent side-payments that they will make after a non-cooperative
strategic form game is played and the payoffs are resolved. These side-payments are binding contracts, so the players are \emph{ex ante} determining their preferences over the game's outcomes. In this regard the game that the players actually play is endogenously determined. JW examine whether a mechanism that allows players to make such outcome-contingent side-payments generally results in efficient outcomes and conclude that it does not. A related idea has been recently explored in the context of mechanism design when considering auctions with aftermarkets \cite{Babaioff2023}.  

The simplest game-manipulation scenario can be viewed as a special case of JW. In this special case, the only outcome-contingent side-payments are between the players and the game manipulators. The game manipulators would be indifferent over the game outcomes if not for the fact that they will be receiving side-payments dependent on those outcomes. Furthermore, the game manipulators play no part in the game between the players other than to accept contracts for outcome-contingent side-payments and make the contracts public. 

In contrast to JW, we do not focus on efficiency issues, and we do not assume that a social planner installs a mechanism for players to make side-payments. Instead, we look at a game without formal mechanisms and ask whether external parties will create contracts for outcome-contingent side-payments in pursuit of profit. In particular, we examine the implications of giving the game manipulator the power to offer contracts, which will in general increase the game manipulator's profits. In addition, we relax the assumption in JW that all side-payments are non-negative. That is, we allow game manipulators to pay players for certain outcomes. This will be important when examining optimal contracts as well as the extent to which a monopolist game manipulator can extract profits from players. We also consider how things change when
there is more than one manipulator, when manipulation contracts are offered in sequence, etc. None
of those issues arise in JW.

In another related paper, \cite{Renou09} analyses what happens when players are able to embed the original game in a new two-stage game. In the second stage of that new game the players play the original game. However before they do so, in the first stage, the players each simultaneously commit not to play some subset of their possible moves in that game in the second stage. These \textit{commitment games} can be seen as another special case of JW in which (1) player $i$'s side-payments are only contingent on $i$'s action (rather than on the full profile of actions), (2) the side-payments are made to external players and (3) the side-payments are effectively infinite. Renou's analysis does not apply to the full game manipulation scenario. This is because there are many circumstances in which both the player and the game manipulator prefer to make contracts that are fully outcome-contingent and that have non-infinite side-payments. One particular example is when the player and game manipulator find it optimal to agree on a contract that results in an equilibrium where the player uses a mixed strategy with full support (and therefore does not make any commitment in the first stage
of Renou's two-stage game).

The idea that there might be pre-game play in which players make choices to affect their own preferences over outcomes is also present in \cite{Wolpert08a}. The authors analyse the idea that experimentally observed non-rationality is in fact rational, because by committing to play the game with a non-rational ``persona,'' a player may increase her ultimate payoff. This persona has the same effect as a side-payment or commitment, as it is reflected in a temporary change to the player's utility function. The \textit{persona games} model has been successful in explaining non-rational behaviour in non-repeated traveler's dilemma and even in versions of the non-repeated prisoner's dilemma.

A related commitment game between the players has been formalised in \cite{Tennenholtz2004} as the notion of program equilibria (see \cite{LaVictoire2014, Oesterheld2019} for further extensions of this model). The program equilibrium model describes a scenario where each player designs a computer program that will determine their actions in the game. As in the persona game, it is necessary that the commitments will be made public: the programs can read the other players' programs, and their actions are contingent on the content of these programs (a simple example of such an equilibrium in the prisoner's dilemma is ``cooperate if the other program is identical to this program or defect otherwise'').

In \cite{KolumbusNisan2022how, KolumbusNisan2022auctions}, the authors study scenarios of repeated games where players are assisted by learning algorithms. It is shown that in auctions and other game classes, the long-term outcomes of learning dynamics induce a meta-game between the players in which the players often have incentives to misreport their true preferences to their own learning algorithms. The commitment that players make in this case, namely, to use learning algorithms, is different in nature from notions of punishments or rewards studied in repeated games, since actions are not contingent on the opponents' play. Interestingly, in this model, the details of the algorithms or the goals provided to them need not be made public, but rather they are revealed through the dynamics.\footnote{Our model naturally extends to repeated games with a fixed contract and to players who use no-regret algorithms, as such algorithms (almost) never play dominated strategies. We focus our discussion, however, on the single-shot case for ease of exposition.}

There is a subset of the principal-agent literature concerning delegation games that is closely related to game manipulation. In these models, the principal is able to contract with an agent that will engage in a game with the principal's opponent (or agent of the principal's opponent). One concern of this literature is detailing the optimal contract for a principal (see \cite{Vickers85,Fershtman85,Sklivas87}). Another concern is whether a mechanism that allows specific types of contracts can lead to Pareto efficiency (see \cite{Kalai91,Katz91}). Game manipulation is closely related to a previously unexplored aspect of principal-agent scenarios: the degree to which the agents can profit from delegation contracts.

Furthermore, our work is related to the general literature on commitment in games because, at its core, game manipulation is about what happens when players benefit by strategically restricting themselves. One well-studied aspect of commitment is the role of timing. Papers such as \cite{Hamilton90}, \cite{vanDamme96} and \cite{Romano05} concern endogenous timing and Stackelberg-like commitments. Another area of study is the role of commitment in repeated games. In their study of finitely repeated games, \cite{Garcia06} introduce a weakening of SPE called virtual subgame perfect equilibrium. \cite{Kalai07} also study commitment in finitely repeated games, but do so in a manner similar to JW. That is, they are concerned with the role of commitment in bringing about efficiency.

The contracts offered by an external party in our case are also related to the broad literature on mechanism design, where a platform offers outcome-contingent incentives to the players, for example, in auctions \cite{Milgrom2004}, capital-raising \cite{Halac2020, Babaioff2022}, and various social choice scenarios \cite{SocialChoice2016}. Perhaps most closely related in this literature is \cite{Monderer2003}, who explore the cost needed from an external designer for implementing (i.e., incentivising) a given outcome in a game. This literature mostly focuses on a mechanism designer who aims to implement socially efficient outcomes. By contrast, we explore the potential implications of a self-interested external entity intervening in the interaction between our players. The latter scenario may, in principle, lead to a negative effect on society as long as the external party has profits.

%%%%%%%%%%%%%%%%%%%%%%%%%%%%%%%%%%%%%%%%%%

\end{document}